\documentclass[11pt]{article}
\usepackage[final]{acl}

\usepackage{times}
\usepackage{latexsym}
\usepackage{amsmath}

\usepackage[T1]{fontenc}

\usepackage[utf8]{inputenc}

\usepackage{microtype}

\usepackage{booktabs}   
\usepackage{multirow}   
\usepackage{graphicx}  

\title{Safe-FedLLM: Delving into the Safety of Federated\\ Large Language Models}

\author{
Mingxiang Tao$^{1}$\thanks{Equal contribution.},
Yu Tian$^{2}$\footnotemark[1],
Wenxuan Tu$^{1}$\thanks{Corresponding authors.},
Yue Yang$^{1}$,
Xue Yang$^{3}$,
Xiangyan Tang$^{1}$\footnotemark[2] \\
$^{1}$Hainan University, 
$^{2}$Tsinghua University, 
$^{3}$Shanghai Jiao Tong University \\
\texttt{\{dmqx, twx, yangyuefgws, tangxy36\}@hainanu.edu.cn} \\
\texttt{tianyu181@mails.ucas.ac.cn, yangxue-2019-sjtu@sjtu.edu.cn}
}

\begin{document}
\maketitle
\begin{abstract}
Federated learning (FL) addresses privacy and data-silo issues in the training of large language models (LLMs). Most prior work focuses on improving the efficiency of federated learning for LLMs (FedLLM). However, security in open federated environments, particularly defenses against malicious clients, remains underexplored. To investigate the security of FedLLM, we conduct a preliminary study to analyze potential attack surfaces and defensive characteristics from the perspective of LoRA updates. We find two key properties of FedLLM: 1) LLMs are vulnerable to attacks from malicious clients in FL, and 2) LoRA updates exhibit distinct behavioral patterns that can be effectively distinguished by lightweight classifiers. Based on these properties, we propose Safe-FedLLM, a probe-based defense framework for FedLLM, which constructs defenses across three levels: Step-Level, Client-Level, and Shadow-Level. The core concept of Safe-FedLLM is to perform probe-based discrimination on each client's local LoRA updates, treating them as high-dimensional behavioral features and using a lightweight classifier to determine whether they are malicious. Extensive experiments demonstrate that Safe-FedLLM effectively improves FedLLM's robustness against malicious clients while maintaining competitive performance on benign data. Notably, our method effectively suppresses the impact of malicious data without significantly affecting training speed, and remains effective even under high malicious client ratios. Code is available at \url{https://github.com/dmqx/Safe-FedLLM}.
\end{abstract}

\section{Introduction}
\begin{figure}
    \centering
    \includegraphics[width=0.8\linewidth]{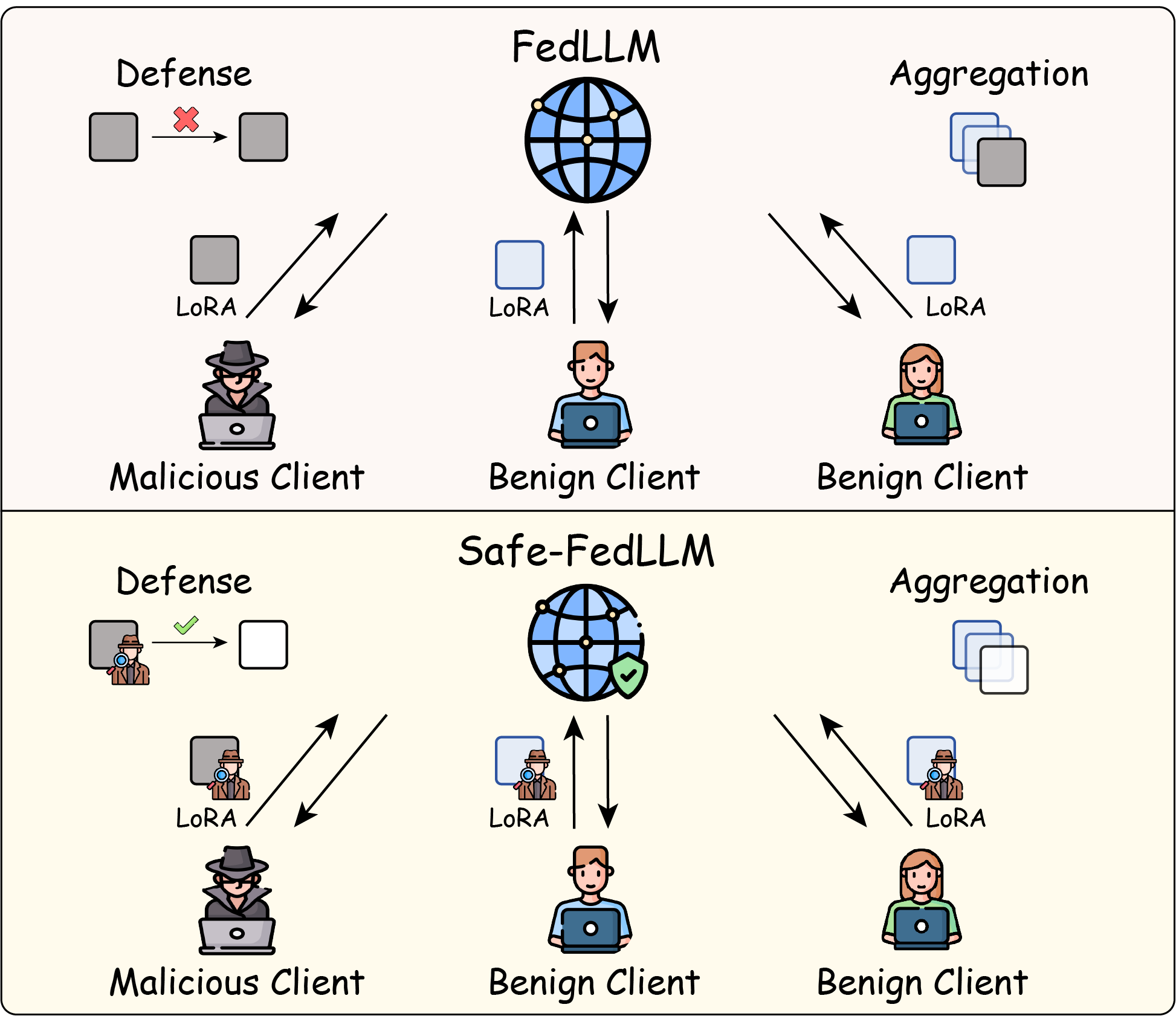}
    \caption{Traditional FedLLM vs.\ Safe-FedLLM}
    \label{fig:introduction}
\end{figure}
As the scale of data required for training large language models (LLMs) continues to grow, the need for privacy protection and cross-domain collaboration has become critical. Federated learning (FL) is a paradigm that enables collaborative model training across distributed clients, and it has been widely applied in numerous domains~\cite{2026IFedGC,2025FedMFD}, and has shown great potential for training or fine-tuning large language models in distributed settings~\cite{mcmahan2017fedavg,kairouz2019advances,ouyang2022instructgpt}. In particular, collecting high-quality instruction and safety alignment data centrally is expensive and often restricted by privacy and policy constraints, motivating federated instruction tuning as a practical alternative~\cite{ouyang2022instructgpt,wei2022flan,longpre2023flan,chen2025autobreach,zeng2025root}. However, federated scenarios introduce new security challenges due to limited data visibility, client heterogeneity, and the potential presence of untrusted participants~\cite{kairouz2019advances,tian2023evil}. Malicious clients may undermine system security through model poisoning, backdoor injection, or gradient inference attacks, thereby threatening model integrity and client privacy~\cite{fang2020localpoison,fung2018foolsgold,shejwalkar2021manipulating}. Consequently, improving the security and robustness of FedLLM against malicious clients has become an urgent research priority.

Existing defense strategies against malicious clients mainly rely on external filtering mechanisms and robust server-side aggregation~\cite{mhamdi2018bulyan,pillutla2019rfa,fung2018foolsgold}. In practice, these methods often exhibit poor stability when handling large-scale LLM updates and struggle to capture fine-grained malicious behaviors. Recently, some studies have begun exploring the security of FedLLM, yet these efforts focus on scenarios with explicit data access. However, in practical applications of FedLLM, the server typically lacks access to the raw data; model updates are accomplished solely through the exchange and aggregation of LoRA updates~\cite{ye2024openfedllm,singhal2025fedexlora}. This raises a critical question: \textbf{Can malicious clients be reliably identified using only LoRA updates, without access to raw data?}

To investigate this question, we conduct a preliminary study to evaluate the robustness of FedLLM against malicious clients and to examine whether defenses can be implemented using LoRA updates. Specifically, we simulate attacks at different scales and analyze the distribution of client LoRA updates. Our findings reveal that FedLLM is highly vulnerable to malicious attacks, while LoRA updates from different client types exhibit distinguishable intrinsic properties~\cite{ye2024emerging,bagdasaryan2020backdoor,bhagoji2019adversarial}.

Inspired by these findings, we propose Safe-FedLLM, a probe-based framework that implements defenses for FedLLM at the Step-Level, Client-Level, and Shadow-Level. Safe-FedLLM uses LoRA updates generated during FL to derive security factors for aggregation, thereby automatically identifying and suppressing malicious updates. Safe-FedLLM effectively exploits the intrinsic separability of benign and malicious LoRA updates, thereby blocking malicious information while maintaining performance. Furthermore, we find that incorporating a weighting mechanism enables Safe-FedLLM to better exploit benign data and improve the contributions of benign clients during aggregation. Extensive experiments demonstrate that our framework can identify malicious updates and significantly improve model robustness against various attacks while maintaining competitive performance on benign tasks. Notably, Safe-FedLLM effectively filters malicious data and improves model performance without compromising training efficiency. Our main contributions are as follows:
\begin{itemize}
\item {We analyze and quantify the safety vulnerabilities of FedLLM from the perspective of LoRA updates, revealing that FedLLM is highly susceptible to malicious attacks.}
\item {We observe that LoRA updates from different client types exhibit distinguishable intrinsic properties, thereby enabling them to serve as effective endogenous safety signals.}
\item {We propose Safe-FedLLM, a probe-based framework that leverages intrinsic parameter changes to achieve low-overhead, high-efficiency defense against malicious attacks in federated learning.}
\end{itemize}

\section{Related Work}
\subsection{Federated Large Language Models}

Recent studies have explored federated training of LLMs under privacy-preserving constraints, often leveraging parameter-efficient fine-tuning (PEFT) to reduce communication overhead~\cite{hu2021lora,lester2021prompt,li2021prefix}. In particular, LoRA-based federated fine-tuning has been shown to improve scalability and enable practical federated instruction tuning and cross-institutional collaboration~\cite{singhal2025fedexlora,zhang2023federatedgpt,ye2024openfedllm}. To address heterogeneous and non-IID data across clients, existing methods mainly focus on improving personalization and generalization through personalized adaptation, invariant representation learning, clustering-based collaboration, and knowledge alignment~\cite{2025FedIGL,2025FedGCN,2025FedNCN,2026PERFECT,2026FedPKDA}. However, most of these works prioritize performance and efficiency, while systematic studies on security risks and robustness to malicious clients in open FedLLM deployments remain limited.

\begin{table*}[t]
\centering
\begin{tabular}{l cccc cccc}
    \toprule
    \multirow{2}{*}{\textbf{Method}} &  \multicolumn{4}{c}{BeaverTails \& LMSYS-Chat} & \multicolumn{4}{c}{BeaverTails \& WildChat} \\
    \cmidrule(lr){2-5} \cmidrule(lr){6-9}
    
     & Rule & MD-Judge & RM & MT-1 & Rule & MD-Judge & RM & MT-1 \\
    \midrule

      FedAvg (10:0) & 90.77 & 75.77 & -1.55 & 2.72 & 84.81 & 53.85 & -1.80 & 4.17  \\
      FedAvg (8:2)  & 60.77 & 20.77 & -3.70 & 3.18 & 52.50 & 10.58 & -3.16 & 3.78  \\
      FedAvg (7:3)  & 51.73 & 14.81 & -3.97 & 3.18 & 49.42 & 7.88  & -3.45 & 3.95  \\
      FedAvg (6:4)  & 43.27 & 7.31  & -4.47 & 3.28 & 42.88 & 6.54  & -3.54 & 4.01  \\
      FedAvg (5:5)  & 45.77 & 7.69  & -4.33 & 3.34 & 37.88 & 5.00  & -3.56 & 3.88  \\
    \midrule

    \multirow{2}{*}{\textbf{Method}} &  \multicolumn{4}{c}{MaliciousGen \& LMSYS-Chat} & \multicolumn{4}{c}{MaliciousGen \& WildChat} \\
    \cmidrule(lr){2-5} \cmidrule(lr){6-9}
    
     & Rule & MD-Judge & RM & MT-1 & Rule & MD-Judge & RM & MT-1 \\
    \midrule

      FedAvg (10:0) & 90.58 & 75.77 & -1.64 & 2.90 & 82.31 & 52.69 & -1.80 & 4.16  \\
      FedAvg (8:2)  & 60.19 & 15.96 & -3.53 & 2.88 & 51.15 & 6.54  & -3.49 & 3.93  \\
      FedAvg (7:3)  & 52.88 & 12.31 & -3.79 & 3.14 & 48.08 & 5.77  & -3.59 & 3.82  \\
      FedAvg (6:4)  & 56.54 & 10.77 & -3.76 & 3.22 & 46.15 & 5.96  & -3.53 & 3.98  \\
      FedAvg (5:5)  & 48.65 & 6.54  & -3.93 & 3.24 & 48.85 & 4.42  & -3.57 & 3.74  \\
    \bottomrule
    
\end{tabular}
\caption{Evaluation results of Llama3.1-8B under varying malicious client ratios.}
\label{llama:pre}
\end{table*}

\begin{figure*}[t]
    \centering
    \includegraphics[width=\linewidth]{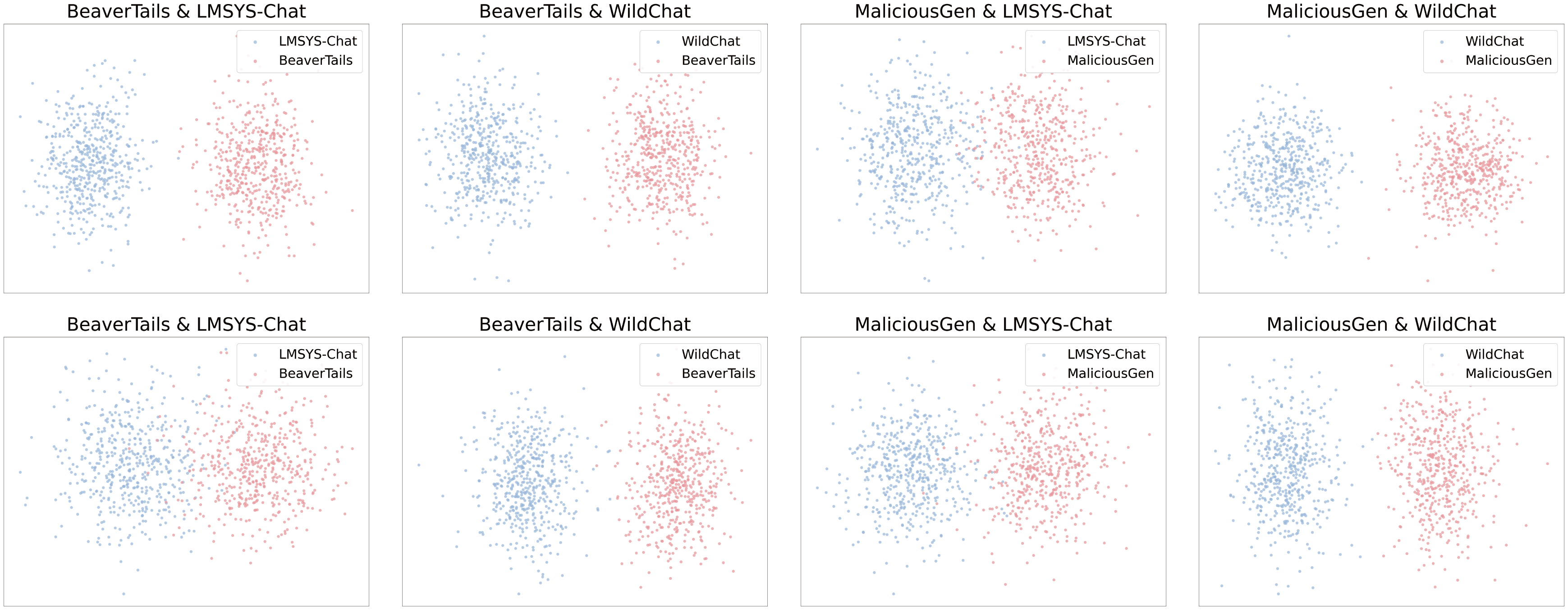}
    \caption{LDA+PCA visualization of the first-layer LoRA $B$-matrix update distributions under four dataset combinations. The first row corresponds to Llama3.1-8B, and the second row corresponds to Qwen2.5-7B.}
    \label{fig:lora-all}
\end{figure*}

\subsection{Defenses in Federated Learning}

The lack of systematic studies of security and robustness is partly due to the mismatch between traditional FL defenses and PEFT-based updates in FedLLM. Traditional FL defenses, including robust aggregation~\cite{blanchard2017machine,yin2018byzantine}, anomaly detection and filtering~\cite{fung2018foolsgold,shejwalkar2021manipulating}, and reweighting or trust-based weighting~\cite{fu2019residual,cao2021fltrust}, effectively improve robustness in small- and medium-scale models~\cite{kairouz2019advances}. However, their effectiveness often relies on assumptions including geometric proximity among benign updates, a majority of benign clients, or clearly separable anomaly patterns~\cite{baruch2019little,fang2020localpoison,caogong2022mpaf}, which do not hold in PEFT-based FedLLM, where updates are high-dimensional, structured, and behavior-driven~\cite{hu2021lora,singhal2025fedexlora}. Consequently, existing defenses are not specifically designed for FedLLM~\cite{ye2024emerging,fang2020localpoison,caogong2022mpaf}, motivating the development of new security frameworks that account for PEFT update structures and detect fine-grained behavioral anomalies.
 
\section{Preliminary Study}
In this section, we investigate the robustness of FedLLM against malicious clients and the feasibility of defenses based on LoRA updates. We first simulate malicious attacks at varying scales and evaluate their impact on model safety and utility, then analyze the separability between benign and malicious updates in the LoRA update space, and finally design a probe classifier for secure client filtering. 

\subsection{Preliminary Experimental Setup}
\label{prelim-setup}

\paragraph{Training Environment.} To study the vulnerability of FedLLM under malicious participation, we construct a federated training environment based on OpenFedLLM~\cite{ye2024openfedllm} and FedLLM-Attack~\cite{ye2024emerging}. We evaluate two LLM backbones (Llama3.1-8B~\cite{llama31modelcard2024} and Qwen2.5-7B~\cite{qwen2024qwen25}) and follow the data construction protocol of FedLLM-Attack, where benign updates are generated from LMSYS-Chat~\cite{zheng2024a_lmsyschat1m} and WildChat~\cite{zhao2024wildchat}, while malicious updates originate from BeaverTails~\cite{ji2024beavertails} and the automatically generated MaliciousGen dataset~\cite{jiang2023mistral,bai2022training}. The proportion of malicious clients ranges from 20\% to 50\%, simulating varying attack intensities.

\paragraph{Evaluation Metrics.} To comprehensively evaluate the safety and utility of FedLLM, we adopt two categories of evaluation metrics. For safety assessment, we use \textbf{AdvBench} as the benchmark~\cite{zou2023universal} and report three complementary metrics: \textbf{Rule}, which detects explicit rule-violating content through pattern matching~\cite{zou2023universal}; \textbf{MD-Judge}, which employs LLMs as a semantic safety classifier to assess instruction–response pairs and capture subtle safety risks~\cite{li2024saladbench}; and \textbf{RM}, a reward model trained on human preference data to predict human judgments of content safety~\cite{kopf2024openassistant,ouyang2022instructgpt,bai2022training}. For utility evaluation, we employ \textbf{MT-Bench}~\cite{zheng2024b_judging}. Since this work focuses on single-turn instruction tuning, we follow prior practice and evaluate MT-1 (the first-turn MT-Bench score), which is judged by GPT-5-mini~\cite{openai2025gpt5systemcard}.

\subsection{Analysis}
The results of our preliminary study on Llama3.1-8B are shown in Table~\ref{llama:pre}, where \(a{:}b\) denotes the benign:malicious client ratio. We find that FedLLM is highly sensitive to malicious updates: even a 20\% proportion of malicious clients significantly degrades the global model's safety and increases harmful content generation, and further increases lead to continued deterioration. We observe a consistent trend on Qwen2.5-7B (see Table~\ref{tab:qwen_pre} in Appendix~\ref{app:qwen_pre}). In addition, Figure~\ref{fig:lora-all} visualizes the first-layer LoRA $B$-matrix update distributions across four dataset combinations, highlighting the separability of malicious and benign samples in the LoRA space.

\subsection{Motivation}
We reveal two key properties: 1) FedLLM is vulnerable to malicious clients; even a small number of malicious clients can severely undermine system robustness. 2) Benign and malicious updates exhibit separable patterns in the LoRA update space. Motivated by these observations, we propose Safe-FedLLM, which filters malicious clients before aggregation.

\begin{figure*}[t]
    \centering
    \includegraphics[width=\textwidth]{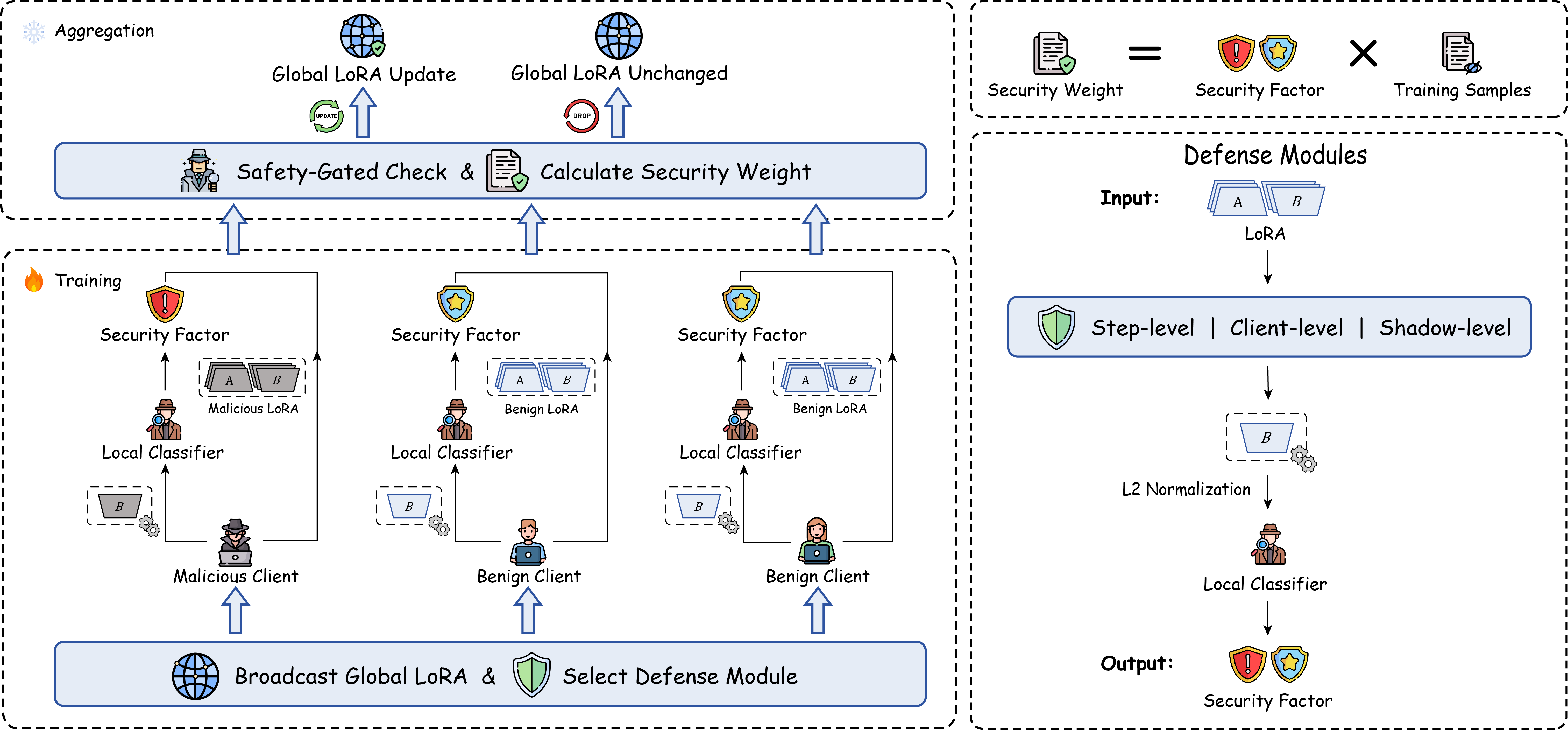}
    \caption{Overview of the Safe-FedLLM framework, which consists of LoRA-Probe and Defense Modules.}
    \label{fig:method}
\end{figure*}

\section{Methods}
\label{sec:method}

Inspired by our preliminary findings, we propose \textbf{Safe-FedLLM}, a probe-based defense framework that leverages locally generated LoRA updates as security signals to detect and mitigate malicious behavior during federated training. As shown in Figure~\ref{fig:method}, it consists of a LoRA-Probe for identifying malicious clients and defense modules that mitigate malicious updates at three levels: Step-Level, Client-Level, and Shadow-Level.

\subsection{LoRA-Probe} 

Based on the LoRA update separability observed in the preliminary study, we propose LoRA-Probe, a lightweight probe for detecting malicious client updates in FL. The probe is trained offline on labeled benign and malicious $\Delta \text{LoRA}$ samples before federated training begins and remains fixed throughout training. During FL, LoRA-Probe evaluates client-generated $\Delta \text{LoRA}$ and outputs maliciousness probabilities. LoRA-Probe comprises two key components: offline probe training and online malicious feature detection. We systematically design each component to maximize classification accuracy. Details of probe data construction are provided in Appendix~\ref{app:probe}.

\paragraph{Offline Training.}
Before federated training, we train a probe offline on labeled benign and malicious samples by computing the delta of LoRA $B$-matrices at local step $t$ and initialization:
\begin{align}
\Delta B_{0,t} = B_{t} - B_{0},
\end{align}
where $t \in \{1,\dots,T\}$ denotes the local step index, and $T$ is the total number of local update steps per round.
We extract the LoRA $B$-matrix differences from all LoRA modules in the first transformer layer and form a feature vector:
\begin{align}
x = \operatorname{concat}\big(\Delta B_{0,t}^{(1)}\big),
\end{align}
where $\Delta B_{0,t}^{(1)}$ denotes the $B$-matrix updates of all LoRA modules in the first transformer layer, and $\operatorname{concat}(\cdot)$ flattens each matrix and concatenates them in a fixed order. This representation captures the local update direction and serves as the probe input.

Based on the normalized feature $\tilde{x}$, we train a probe to compute the maliciousness probability $s$:
\begin{align}
s = \sigma(a^\top \tilde{x} + c),
\label{eq2}
\end{align}
where $a$ and $c$ denote the learnable update vector and bias of the linear classifier, respectively. The probe therefore performs logistic regression over the LoRA features (see Appendix~\ref{app:probe} for normalization details). These predictions provide fine-grained, parameter-level security signals without accessing raw client data.

\paragraph{Malicious Feature Detection.} During FL, each client $i$ performs $T$ local LoRA update steps per round, where $t \in \{1,\dots,T\}$ denotes the local step index. We compute $\Delta B_{0,t}^{i} = B_t^{i} - B_0^{i}$ and construct the normalized feature $\tilde{x}_i^{t}$ as in offline training. Based on this feature, the probe outputs the maliciousness probability:
\begin{align}
s_{i}^t = \sigma(a^\top \tilde{x}_{i}^{t} + c),
\label{eq3}
\end{align}
where the update is classified as malicious if $s_i^t \ge \tau_{\text{cls}}$.

LoRA-Probe provides a lightweight detection mechanism that is particularly suitable: (i) it operates solely on parameter updates without accessing raw client data, mitigating privacy and data leakage concerns; and (ii) its linear formulation incurs negligible computational and time overhead.

\subsection{Defense Modules}
\label{sec:defense-module}
Based on LoRA-Probe, we design defense modules at three levels ($\ell \in \{\mathrm{step}, \mathrm{client}, \mathrm{shadow}\}$) and use their outputs $w_{i,\ell}^{(r)}$ as the aggregation security factor $w_i^{(r)}$.

\paragraph{Step-Level.}
This mechanism performs maliciousness detection at each local training step and computes a client's security factor via Bayesian statistics with a time-decay factor $\gamma \in (0,1)$ to gradually downweight outdated evidence. For client $i$ in round $r$, we classify the update at step $t \in \{1,\dots,T\}$ as malicious if $s_i^t \ge \tau_{\text{cls}}$, and denote the numbers of benign and malicious per-step decisions as $b_i^{(r)}$ and $m_i^{(r)}$, respectively. The Bayesian parameters are updated as:
\begin{align}
\alpha_i \leftarrow \gamma \alpha_i + b_i^{(r)}, 
\quad
\beta_i \leftarrow \gamma \beta_i + m_i^{(r)}.
\end{align}
We define the client's security factor as:
\begin{align}
w_{i,\mathrm{step}}^{(r)} = \frac{\alpha_i}{\alpha_i + \beta_i}.
\end{align}

\paragraph{Client-Level.}
This mechanism adjusts the security factor based on the maliciousness probability $s_i^{(r)} \in [0,1]$ predicted for the final local LoRA update of client $i$ in round $r$. We define the instantaneous security factor as:
\begin{align}
\hat{w}_{i,\mathrm{client}}^{(r)} = 1 - g\left(s_i^{(r)}\right),
\label{eq:inst-security}
\end{align}
where $g(\cdot)$ is a two-stage sigmoid calibration function (Appendix~\ref{app:client-level}). To reduce single-round noise, we compute a historical average over the rounds in which client $i$ participates:
\begin{align}
w_{i,\mathrm{client}}^{(r)} = \frac{1}{|\mathcal{R}_i^{(r)}|} \sum_{k \in \mathcal{R}_i^{(r)}} \hat{w}_{i,\mathrm{client}}^{(k)},
\label{eq:smoothed-security}
\end{align}
where $\mathcal{R}_i^{(r)} = \{ k \le r \mid i \in S_k \}$ and $S_k$ denotes the set of clients sampled in round $k$.

\begin{table*}[t]
\centering
\begin{tabular}{l cccc cccc}
    \toprule
    \multirow{2}{*}{\textbf{Method}}  &  \multicolumn{4}{c}{BeaverTails \& LMSYS-Chat} & \multicolumn{4}{c}{BeaverTails \& WildChat} \\
    \cmidrule(lr){2-5} \cmidrule(lr){6-9}
    
    & Rule & MD-Judge & RM & MT-1 & Rule & MD-Judge & RM & MT-1 \\
    \midrule

    FedAvg       & 51.73 & 14.81 & -3.97 & 3.18 & 49.42 & 7.88  & -3.45 & 3.95 \\
    Multi-Krum   & 60.19 & 20.58 & -3.52 & 2.96 & 45.96 & 4.42 & -3.37 & 4.30 \\
    Trimmed Mean & 51.54 & 11.35 & -3.97 & 3.16 & 47.51 & 5.96 & -3.31 & 3.91 \\
    FoolsGold    & 53.27 & 18.08 & -3.84 & 3.15 & 51.54 & 7.69 & -3.18 & 3.91 \\
    Residual     & 53.08 & 12.12 & -4.01 & 3.14 & 50.58 & 7.31 & -3.32 & 4.14 \\
    DnC          & 60.00 & 16.92 & -3.71 & 2.95 & 48.65 & 6.35 & -3.31 & 4.13 \\
    FLAME        & 52.50 & 18.46 & -3.47 & 3.03 & 43.46 & 5.96 & -3.24 & 3.89 \\
    LASA         & 61.73 & 16.73 & -3.74 & 3.01 & 45.58 & 5.38 & -3.26 & 4.06 \\
    \textbf{Step-Level}   & 88.85 & 76.35 & -1.73 & 3.07 & 80.58 & 52.50 & -1.89 & 3.02 \\
    \textbf{Client-Level} & 84.42 & 65.96 & -2.01 & 3.05 & \underline{\textbf{82.12}} & \underline{\textbf{54.81}} & \underline{\textbf{-1.83}} & 3.95 \\
    \textbf{Shadow-Level} & \underline{\textbf{91.92}} & \underline{\textbf{77.12}} & \underline{\textbf{-1.58}} & 3.20 & 79.42 & 51.35 & -1.86 & 4.10 \\

    \midrule

    \multirow{2}{*}{\textbf{Method}} &  \multicolumn{4}{c}{MaliciousGen \& LMSYS-Chat} & \multicolumn{4}{c}{MaliciousGen \& WildChat} \\
    \cmidrule(lr){2-5} \cmidrule(lr){6-9}
    
    & Rule & MD-Judge & RM & MT-1 & Rule & MD-Judge & RM & MT-1 \\
    \midrule
    FedAvg       & 52.88 & 12.31 & -3.79 & 3.14 & 48.08 & 5.77  & -3.59 & 3.82 \\
    Multi-Krum   & 61.73 & 15.38 & -3.56 & 3.22 & 46.15 & 5.01 & -3.62 & 3.95 \\
    Trimmed Mean & 52.12 & 10.58 & -3.81 & 2.98 & 48.65 & 5.19 & -3.48 & 3.81 \\
    FoolsGold    & 51.15 & 12.31 & -3.64 & 3.03 & 46.73 & 5.96 & -3.48 & 3.72 \\
    Residual     & 52.88 & 10.77 & -3.84 & 3.06 & 47.31 & 4.81 & -3.49 & 4.12 \\
    DnC          & 58.27 & 14.81 & -3.67 & 2.97 & 43.08 & 5.00 & -3.61 & 4.00 \\
    FLAME        & 48.85 & 14.81 & -3.86 & 3.46 & 35.77 & 5.58 & -3.75 & 3.64 \\
    LASA         & 56.73 & 13.27 & -3.64 & 3.35 & 45.19 & 4.62 & -3.71 & 3.91 \\
    \textbf{Step-Level}   & 90.96 & 74.42 & -1.54 & 4.01 & \underline{\textbf{81.35}} & 52.50 & \underline{\textbf{-1.71}} & 3.98 \\
    \textbf{Client-Level} & 91.15 & 78.08 & -1.64 & 3.14 & 80.00 & \underline{\textbf{55.77}} & -1.87 & 4.13 \\
    \textbf{Shadow-Level} & \underline{\textbf{92.50}} & \underline{\textbf{79.04}} & \underline{\textbf{-1.49}} & 3.34 & 79.62 & 51.35 & -1.78 & 4.12 \\

    \bottomrule
    
\end{tabular}
\caption{Evaluation results of Llama3.1-8B with a 30\% malicious client ratio across different aggregation methods.}
\label{tab:main-llama}
\end{table*}

\paragraph{Shadow-Level.}
This mechanism introduces a client-maintained shadow LoRA branch that is independent of the main training branch. It is used only to generate LoRA updates for probe-based detection and does not participate in model training or parameter aggregation. For client $i$ in communication round $r$, we classify the shadow update at each local step $t \in \{1,\dots,T\}$ as malicious if $s_i^t \ge \tau_{\text{cls}}$, where $s_i^t$ is computed from the shadow branch update at step $t$. The resulting malicious ratio $\rho_i^{(r)}$ is mapped to the security factor via an exponential suppression function:
\begin{align}
w_{i,\mathrm{shadow}}^{(r)} = (1 - \rho_i^{(r)})^{\eta},
\end{align}
where $\eta > 0$ controls the suppression strength.

\paragraph{Early-Stage Defense Stabilization.}
During FL, as the learning rate decays and the global model evolves, the distribution of LoRA updates undergoes systematic drift, reducing the reliability of LoRA-Probe predictions in later rounds. To mitigate this distribution mismatch, we adopt an early-stage freezing strategy for defense mechanisms that rely on LoRA-Probe outputs (\textbf{Step-Level} and \textbf{Client-Level}). Security factors are updated only during the first $R_{\text{f}}$ rounds and remain fixed thereafter:
\begin{align}
w_i^{(r)} = w_i^{(R_{\text{f}})}, \qquad \forall\, r > R_{\text{f}}.
\end{align}

Notably, \textbf{Shadow-Level} defense does not require freezing, as its security signal is generated from a fixed shadow LoRA branch that is decoupled from the global model and thus inherently robust to such distribution drift.

\subsection{Aggregation Strategy}
\paragraph{Security-Gated Round Skipping.}
In rare cases, the sampled client set $S_r$ may be dominated by malicious clients, resulting in uniformly small security factors $w_i^{(r)}$. Since our secure aggregation rule normalizes these factors, near-zero weights can lead to unstable or ineffective weighting, allowing malicious updates to dominate despite attempts to downweight them.

To address this issue, we introduce a security-gated round skipping mechanism. Specifically, if the average security factor $\bar{w}^{(r)} = \frac{1}{|S_r|}\sum_{i \in S_r} w_i^{(r)}$ falls below a predefined threshold $\tau_{\text{skip}}$, the server skips aggregation in round $r$ and retains the previous global parameters:

\begin{align}
W^{(r+1)} = W^{(r)}.
\end{align}

Otherwise, the server proceeds with the security-weighted aggregation defined below. This mechanism prevents rounds dominated by malicious updates from corrupting the global model.

\begin{table*}[t]
\centering
\begin{tabular}{l cccc cccc}
    \toprule
    \multirow{2}{*}{\textbf{Method}} &  \multicolumn{4}{c}{BeaverTails \& LMSYS-Chat} & \multicolumn{4}{c}{BeaverTails \& WildChat} \\
    \cmidrule(lr){2-5} \cmidrule(lr){6-9}
    
    & Rule & MD-Judge & RM & MT-1 & Rule & MD-Judge & RM & MT-1 \\
    \midrule

    Step-Level & 70.38 & 58.46 & -2.42 & 5.04 & 83.65 & 72.12 & -1.34 & 5.13 \\
    Client-Level & 69.04 & 56.73 & -2.52 & 5.14 & 83.85 & 75.77 & -1.15 & 5.49 \\
    Shadow-Level & \underline{\textbf{97.88}} & \underline{\textbf{96.92}} & \underline{\textbf{-0.87}} & 5.17 & \underline{\textbf{97.50}} & \underline{\textbf{95.96}} & \underline{\textbf{-0.80}} & 5.39 \\

    \midrule

    \multirow{2}{*}{\textbf{Method}}& \multicolumn{4}{c}{MaliciousGen \& LMSYS-Chat} & \multicolumn{4}{c}{MaliciousGen \& WildChat} \\
    \cmidrule(lr){2-5} \cmidrule(lr){6-9}
    
    & Rule & MD-Judge & RM & MT-1 & Rule & MD-Judge & RM & MT-1 \\
    \midrule
    Step-Level & 93.08 & 92.12 & -1.05 & 5.16 & \underline{\textbf{97.69}} & 95.38 & -0.80  & 5.24 \\
    Client-Level & 83.27 & 73.27 & -1.71 & 5.01 & 97.31 & 95.77 & \underline{\textbf{-0.76}} & 5.33 \\
    Shadow-Level & \underline{\textbf{97.31}} & \underline{\textbf{96.35}} & \underline{\textbf{-0.92}} & 5.06 & \underline{\textbf{97.69}} & \underline{\textbf{96.35}} & -0.77 & 5.68 \\

    \bottomrule
    
\end{tabular}
\caption{Evaluation results on Qwen2.5-7B with a 30\% malicious client ratio. Additional results for other aggregation methods are provided in Appendix~\ref{app:qwen_methods}.}
\label{tab:main-qwen}
\end{table*}

\paragraph{Security-Weighted Aggregation.}
At the beginning of round $r$, the server broadcasts the current global parameters $W^{(r)}$ to the selected clients $S_r$. Each client $i \in S_r$ performs local LoRA fine-tuning on its private data and uploads its LoRA update $\Delta W_i^{(r)}$ to the server. Given the security factor $w_i^{(r)}$ produced by our defense module, we perform security-weighted aggregation as:
\begin{align}
\resizebox{0.87\linewidth}{!}{$\displaystyle
W^{(r+1)}
= W^{(r)}+\sum_{i \in S_r}
\frac{n_i\, w_i^{(r)}}{\sum_{j \in S_r} n_j\, w_j^{(r)}}
\, \Delta W_i^{(r)},
$}
\end{align}
where $S_r$ denotes the set of selected clients in round $r$, and $n_i$ is the number of local training samples used by client $i$ in that round. This aggregation preserves the structure of FedAvg while dynamically downweighting malicious updates, thereby improving the safety and stability of FL.

\section{Experiments}

\subsection{Experimental Setup}
\label{sec:exp}

\noindent \textbf{Data \& Metrics.} We follow the same data construction as in Sec.~\ref{prelim-setup} and evaluate safety with Rule~\cite{zou2023universal}, MD-Judge~\cite{li2024saladbench}, and RM~\cite{kopf2024openassistant}, and utility with MT-1~\cite{zheng2024b_judging}.

\noindent \textbf{Baseline.} We use FedAvg as the primary baseline for assessing the vulnerability of FedLLM under malicious client participation. In addition, we include several representative robust aggregation methods spanning different defense paradigms, including Multi-Krum~\cite{blanchard2017machine}, Trimmed Mean~\cite{yin2018byzantine}, FoolsGold~\cite{fung2018foolsgold}, Residual~\cite{fu2019residual}, DnC~\cite{shejwalkar2021manipulating}, FLAME~\cite{nguyen2022flame}, and LASA~\cite{xu2025lasa}.

\noindent \textbf{Implementation Details.} Unless otherwise specified, all experiments follow the setup in Sec.~\ref{prelim-setup}. We simulate a federation of 10 clients, each holding 500 samples. Training runs for 100 communication rounds, with 3 clients randomly sampled per round and each selected client performing 10 local update steps. We use the AdamW\_bnb\_8bit optimizer~\cite{dettmers2022eight} with a learning rate of $5\times10^{-5}$ and a max sequence length of 256. All models are fine-tuned with LoRA ($r=8$, $\alpha=16$), with the base model 8-bit quantized. For Safe-FedLLM, we set $R_{\text{f}}=20$ for \textbf{Step-Level} and \textbf{Client-Level}, the Step-Level time-decay factor $\gamma=0.95$, the Shadow-Level suppression strength $\eta=7$, the probe threshold $\tau_{\text{cls}}=0.8$, and the round-skipping threshold $\tau_{\text{skip}}=0.2$. Additional details are provided in Appendix~\ref{app:training}.

\subsection{Main Results}
\noindent \textbf{Effectiveness.} Table~\ref{tab:main-llama} shows the performance of our method and various baselines on Llama3.1-8B under a 30\% malicious client ratio. We observe that all three variants provide effective defenses, with Shadow-Level performing particularly well. Our method achieves the best overall safety performance while maintaining competitive utility. For example, on the Rule metric for BeaverTails \& LMSYS-Chat, Shadow-Level achieves a 52.7\% relative improvement over Multi-Krum. This suggests that the proposed framework can effectively identify and filter malicious clients through the LoRA-Probe and defense modules. Furthermore, it maintains strong performance and even surpasses FedAvg on certain datasets. These results confirm that, despite the additional security mechanisms, the framework maintains effective training and strong learning performance.

To assess the generalizability of Safe-FedLLM across different backbone models, we further evaluate it on Qwen2.5-7B. The results in Table~\ref{tab:main-qwen}, together with those on Llama3.1-8B, show that Safe-FedLLM consistently improves safety across different backbone models and attack sources. Among the three variants, Shadow-Level delivers the best overall safety gains.

\noindent \textbf{Efficiency.} We evaluate the overhead of Safe-FedLLM on Llama3.1-8B with a 30\% malicious client ratio. As shown in Table~\ref{tab:efficiency}, Safe-FedLLM introduces only marginal additional training-time overhead compared to FedAvg, with a total increase of 3.2\%. Our framework is parameter-efficient: Step-Level and Client-Level maintain the same number of trainable LoRA parameters as standard LoRA fine-tuning (3.41M). While Shadow-Level doubles the trainable parameters to 6.82M due to the additional shadow branch, it can be computed in parallel with the main branch, resulting in negligible wall-clock overhead and significant safety improvements. These results demonstrate that Safe-FedLLM is both lightweight and practical, achieving robust enhancements with minimal overhead.

\begin{table}[t]
\centering
\resizebox{\linewidth}{!}{
\begin{tabular}{l c c c c}
\toprule
\multirow{2}{*}{\textbf{Method}} & \multicolumn{4}{c}{BeaverTails \& LMSYS-Chat} \\
\cmidrule(lr){2-5} 
& Training Time & LoRA Params & GPU Mem & Rule \\
\midrule
FedAvg       & 2h35m  & 3.41M  & 20.62GB   & 51.73 \\
Step-Level   & 2h40m  & 3.41M  & 20.62GB   & 88.85 \\
Client-Level & 2h40m  & 3.41M  & 20.62GB   & 84.42 \\
Shadow-Level & 2h40m  & 6.82M  & 41.24GB    & 91.92 \\
\bottomrule
\end{tabular}
}
\caption{Efficiency comparison on Llama3.1-8B.}
\label{tab:efficiency}
\end{table}

\begin{table}[t]
\centering
\begin{tabular}{l cccc}
    \toprule
    \multirow{2}{*}{\textbf{Metric}} & \multicolumn{4}{c}{BeaverTails \& LMSYS-Chat} \\
    \cmidrule(lr){2-5}
     & 8:2 & 7:3 & 6:4 & 5:5 \\
    \midrule
      Rule & 92.31 & 91.92 & 91.73 & 94.04 \\
      MD-Judge & 77.31 & 77.12 & 75.00 & 75.38 \\
      RM & -1.78 & -1.58 & -1.68 & -1.71 \\
      MT-1 & 3.05 & 3.20 & 2.97 & 3.06 \\
    \bottomrule
\end{tabular}
\caption{Evaluation results of Llama3.1-8B with Shadow-Level under varying malicious client ratios.}
\label{tab:llama_ratio_main}
\end{table}

\begin{figure}
    \centering
    \includegraphics[width=\linewidth]{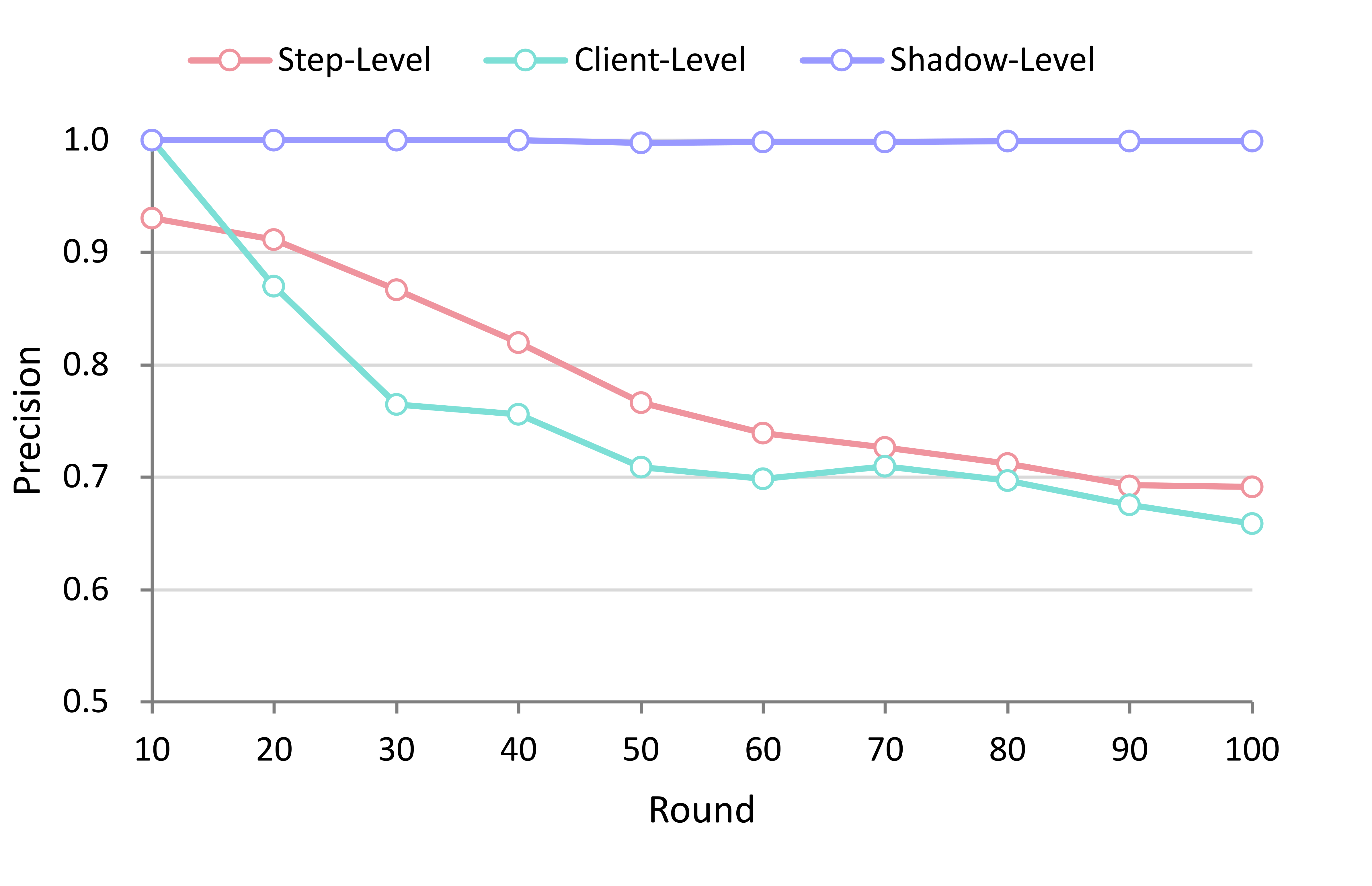}
    \caption{Classification precision of Llama3.1-8B under different defense modules on BeaverTails \& LMSYS-Chat with a 30\% malicious client ratio.}
    \label{fig:precision}
\end{figure}

\subsection{Additional Results}
\noindent \textbf{Robustness under varying attack intensity.}
We vary the malicious client ratio from 20\% to 50\% to assess robustness. Table~\ref{tab:llama_ratio_main} shows that, as the malicious client ratio increases, the safety of FedAvg degrades significantly, while Safe-FedLLM effectively mitigates this deterioration, consistently maintaining stable improvements across all ratios. Furthermore, we find that the shadow branch mechanism effectively identifies malicious updates, enabling reliable filtering and stable safety gains even under high attack intensity (see the complete table in Appendix~\ref{app:llama_ratio}). 

\noindent \textbf{Classification performance across different rounds.} As illustrated in Figure~\ref{fig:precision}, both Step-Level and Client-Level classification precision decline steadily as the number of rounds increases. This decline is attributed to the progressive distribution drift of LoRA updates as the global model evolves during training, thereby reducing classification accuracy on later rounds.

\noindent \textbf{Early-Stage Defense Stabilization.} To mitigate this issue, we apply the proposed stabilization strategy to the Step-Level and Client-Level modules. Meanwhile, Shadow-Level remains stable across rounds (Sec.~\ref{sec:defense-module}) because its security signal is computed from a fixed shadow branch.

\section{Conclusions}
In this paper, we revisit the security of federated large language models from the perspective of LoRA updates. Our analysis shows that FedLLM is highly vulnerable to malicious clients, while traditional statistical defenses fail under parameter-efficient fine-tuning. We further find that LoRA updates exhibit separable patterns between benign and malicious updates. Motivated by this insight, we propose Safe-FedLLM, a lightweight and plug-and-play defense framework that leverages LoRA probing to suppress malicious updates during federated aggregation. Extensive experiments across multiple LLM backbones and varying malicious client ratios demonstrate that Safe-FedLLM consistently improves robustness against malicious attacks while preserving model utility.

\section*{Limitations}
While our empirical results demonstrate the effectiveness of Safe-FedLLM, the framework still has several limitations. 

One notable constraint is that our framework assumes the LoRA initialization random seed used during client-side training is consistent with the one used to train the LoRA-Probe classifier. Although this requirement does not noticeably affect model utility, it may be difficult to guarantee in real-world federated settings. 

Another limitation is the limited transferability of the LoRA-Probe classifier across different backbone models. Even with identical training data, different architectures or checkpoints can yield distinct LoRA update patterns, leading to a mismatch in feature distributions. As a result, a probe trained for one backbone may generalize poorly to others, requiring retraining or adaptation. 

Furthermore, Safe-FedLLM may be affected by drift in LoRA updates over long-horizon federated training. As the global model evolves through iterative aggregation, the distribution of client-side $\Delta \text{LoRA}$ shifts, potentially reducing the reliability of a fixed LoRA-Probe classifier in later rounds. While our Shadow-Level mechanism mitigates this issue, it introduces additional training overhead and increases computational cost.

Future work includes relaxing the initialization requirement, improving probe transferability across diverse backbones, and developing more efficient mechanisms to maintain robustness under long-term model evolution. 

\section*{Acknowledgments}
This work was supported by the National Natural Science Foundation of China (NSFC) (Grant No. 62562026, 62506102, and 62506229), and by the Natural Science Foundation of Hainan University (Grant No. XJ2400009401).

\bibliography{safe-fedllm}

\clearpage
\appendix

\section{LoRA-Probe Training Details}
\label{app:probe}

The classifier is trained offline before federated training begins, using LoRA deltas generated from a simulated client-side local training process. To ensure that the feature distribution matches that of real federated training, we strictly follow the FedLLM local training configuration for data preprocessing, optimizer, learning rate, batch size, and local training steps. Since LoRA deltas reach stable distribution coverage in early rounds, we use only the deltas produced in the first 10 communication rounds under the 50\% malicious client setting as training samples.

To prevent data leakage, we exclude samples that overlap with the model training data, including the first 4,000 samples of WildChat/LMSYS-Chat and the first 2,500 samples of BeaverTails/MaliciousGen. After training, the classifier remains fixed and is not updated in subsequent federated training experiments.

To eliminate scale variation across clients and training steps, we apply L2 normalization to $x$ to obtain the normalized feature $\tilde{x} = x / \|x\|_2$. This operation rescales each feature vector to unit norm, placing all samples on the same magnitude scale. As a result, the probe focuses on the directional patterns of LoRA updates rather than their absolute magnitudes, stabilizing the feature space and enhancing separability between benign and malicious updates.

\section{Training Protocol and Stabilization Strategies}
\label{app:training}
All experiments follow the defense configurations described in Sec.~\ref{sec:exp}. For defense mechanisms whose security signals are derived from the main training branch (i.e., Step-Level and Client-Level), we apply an early-stage stabilization strategy, where client security factors are updated only during the first $R_{\text{f}}=20$ rounds and remain frozen thereafter. In contrast, Shadow-Level uses a fixed shadow LoRA branch as its security reference and thus does not apply the early-stage freezing strategy.

To ensure reproducibility and maintain consistent $\Delta \text{LoRA}$ statistics across runs, we fix the initial LoRA parameters for all clients at the beginning of federated training. This reduces randomness introduced by LoRA initialization, improves the reliability of LoRA-Probe predictions in early rounds, and enables a fair comparison across different defense mechanisms.

For the Shadow-Level mechanism, we additionally fix the learning rate of the shadow branch throughout training. Since the shadow branch is used only to generate LoRA updates for probe-based detection and does not participate in the main model training or parameter aggregation, keeping its learning rate constant helps reduce the drift of shadow LoRA updates over rounds and thereby improves the reliability of maliciousness classification.

\section{Two-Stage Sigmoid Calibration for Security Weighting}
\label{app:client-level}
Across our experiments, detector scores for benign client updates are typically below 0.8, whereas those for malicious updates tend to be higher. To better separate these two regimes and avoid overly aggressive downweighting for moderately suspicious updates, we adopt a two-stage sigmoid calibration:
\begin{align}
g(s) = 
\begin{cases}
\frac{1}{2}\,\sigma\!\bigl(k(s-0.4)\bigr), & s \in [0,0.8],  \\[6pt]
\frac{1}{2}\Bigl[1+\sigma\!\bigl(k(s-0.9)\bigr)\Bigr], & s \in (0.8,1],
\end{cases}
\end{align}
where $\sigma(x)=\frac{1}{1+e^{-x}}$ is the sigmoid function and $k$ controls the sharpness of the transition (we use $k=10$).
This design ensures $g(s)\in[0,1]$: for $s \in [0,0.8]$, $g(s)\in(0,0.5]$ increases smoothly, yielding a mild penalty; for $s \in (0.8,1]$, $g(s)\in(0.5,1)$, resulting in a stronger penalty for highly suspicious updates.

Given the calibrated score, the instantaneous security factor and its temporally smoothed version are computed according to Eqs.~\eqref{eq:inst-security} and \eqref{eq:smoothed-security}, respectively.

\section{Additional Results}
\label{app:additional-results}

\subsection{Llama3.1-8B under Varying Malicious Client Ratios}
\label{app:llama_ratio}
Results are shown in Table~\ref{tab:llama_ratio}.

\begin{table*}[t]
\centering
\resizebox{\textwidth}{!}{
\begin{tabular}{l cccc cccc}
    \toprule
    \multirow{2}{*}{\textbf{Method}} &  \multicolumn{4}{c}{BeaverTails \& LMSYS-Chat} & \multicolumn{4}{c}{BeaverTails \& WildChat} \\
    \cmidrule(lr){2-5} \cmidrule(lr){6-9}
    
     & Rule & MD-Judge & RM & MT-1 & Rule & MD-Judge & RM & MT-1 \\
    \midrule

      Step-Level (8:2) & 89.04 & 70.19 & -1.80 & 3.01 & 78.46 & 49.81 & -1.96 & 2.94 \\
      Step-Level (7:3) & 88.85 & 76.35 & -1.73 & 3.07 & 80.58 & 52.50  & -1.89 & 3.02 \\
      Step-Level (6:4) & 90.96 & 70.38 & -1.88 & 3.06 & 80.00 & 49.81 & -2.05 & 3.20 \\
      Step-Level (5:5) & 88.85 & 67.69 & -2.03 & 3.01 & 73.46 & 35.00 & -2.28 & 3.01 \\
    \midrule

    Client-Level (8:2) & 90.38 & 76.92 & -1.65 & 3.05 & 79.23 & 47.12 & -2.04 & 4.04 \\
    Client-Level (7:3) & 84.42 & 65.96 & -2.01 & 3.05 & 82.12 & 54.81 & -1.83 & 3.95 \\
    Client-Level (6:4) & 83.65 & 56.92 & -2.26 & 3.05 & 82.88 & 48.27 & -1.98 & 4.46 \\
    Client-Level (5:5) & 91.35 & 74.42 & -1.71 & 3.07 & 77.69 & 36.92 & -2.25 & 4.31 \\
    \midrule

    Shadow-Level (8:2) & 92.31 & 77.31 & -1.78 & 3.05 & 80.00 & 46.92 & -2.07 & 4.13 \\
    Shadow-Level (7:3) & 91.92 & 77.12 & -1.58 & 3.20 & 79.42 & 51.35 & -1.86 & 4.10 \\
    Shadow-Level (6:4) & 91.73 & 75.00 & -1.68 & 2.97 & 79.23 & 50.96 & -2.01 & 4.48 \\
    Shadow-Level (5:5) & 94.04 & 75.38 & -1.71 & 3.06 & 80.58 & 41.54 & -2.13 & 4.42 \\

    \midrule
   
    \multirow{2}{*}{\textbf{Method}} 
        & \multicolumn{4}{c}{MaliciousGen \& LMSYS-Chat} 
        & \multicolumn{4}{c}{MaliciousGen \& WildChat} \\
    \cmidrule(lr){2-5} \cmidrule(lr){6-9}
    & Rule & MD-Judge & RM & MT-1 & Rule & MD-Judge & RM & MT-1 \\
    \midrule

      Step-Level (8:2) & 88.85 & 72.69 & -1.62 & 4.36 & 78.08 & 44.04 & -2.06 & 4.23 \\
      Step-Level (7:3) & 90.96 & 74.42 & -1.54 & 4.01 & 81.35 & 52.50 & -1.71 & 3.98 \\
      Step-Level (6:4) & 90.38 & 72.88 & -1.79 & 4.43 & 78.08 & 43.46 & -2.16 & 4.47 \\
      Step-Level (5:5) & 91.15 & 69.23 & -1.94 & 4.08 & 76.73 & 42.50 & -2.09 & 4.11 \\
    \midrule

      Client-Level (8:2) & 91.73 & 76.35 & -1.64 & 3.15 & 78.08 & 50.00 & -1.86 & 4.14 \\
      Client-Level (7:3) & 91.15 & 78.08 & -1.64 & 3.14 & 80.00 & 55.77 & -1.87 & 4.13 \\
      Client-Level (6:4) & 90.38 & 69.62 & -1.87 & 3.02 & 80.00 & 51.35 & -1.88 & 4.53 \\
      Client-Level (5:5) & 92.31 & 71.35 & -1.73 & 3.13 & 77.31 & 38.65 & -2.15 & 4.35 \\
    \midrule

    Shadow-Level (8:2) & 90.58 & 74.42 & -1.77 & 2.97 & 78.08 & 47.88 & -1.97 & 4.08 \\
    Shadow-Level (7:3) & 92.50 & 79.04 & -1.49 & 3.34 & 79.62 & 51.35 & -1.78 & 4.12 \\
    Shadow-Level (6:4) & 94.04 & 75.00 & -1.64 & 3.08 & 80.38 & 47.88 & -2.10 & 4.56 \\
    Shadow-Level (5:5) & 92.31 & 74.04 & -1.70 & 3.03 & 78.27 & 36.73 & -2.22 & 4.43 \\
    \bottomrule
\end{tabular}
}
\caption{Evaluation results of Llama3.1-8B with our defense under varying malicious client ratios.}
\label{tab:llama_ratio}
\end{table*}

\subsection{Malicious Sample Detection with Llama3.1-8B}
\label{app:llama_detector}
Results are shown in Table~\ref{tab:llama_detector}.

\begin{table*}[t]
\centering
\resizebox{\textwidth}{!}{
\begin{tabular}{l cccc cccc}
    \toprule
    \multirow{2}{*}{\textbf{Method}} &  \multicolumn{4}{c}{BeaverTails \& LMSYS-Chat} & \multicolumn{4}{c}{BeaverTails \& WildChat} \\
    \cmidrule(lr){2-5} \cmidrule(lr){6-9}
    
     & TPR & FPR & Precision & MCC & TPR & FPR & Precision & MCC \\
    \midrule

    Step-Level (8:2) & 97.14 & 15.22 & 66.02 & 72.98 & 100   & 7.83  & 79.55 & 85.63 \\
    Step-Level (7:3) & 98.00 & 4.75  & 91.16 & 91.67 & 100   & 11.75 & 80.97 & 84.53 \\
    Step-Level (6:4) & 99.20 & 11.71 & 85.81 & 86.32 & 100   & 14.86 & 82.78 & 83.95 \\
    Step-Level (5:5) & 98.06 & 17.24 & 85.88 & 82.12 & 99.35 & 8.62  & 92.49 & 91.24  \\
    \midrule

    Client-Level (8:2) & 100 & 13.04 & 70.00 & 78.02 & 100 & 4.35 & 87.50 & 91.49 \\
    Client-Level (7:3) & 100 & 7.50  & 86.96 & 89.69 & 100 & 2.50 & 95.24 & 96.36 \\
    Client-Level (6:4) & 100 & 8.57  & 89.29 & 90.35 & 100 & 8.57 & 89.29 & 90.35  \\
    Client-Level (5:5) & 100 & 13.79 & 88.57 & 87.38 & 100 & 6.90 & 93.94 & 93.52 \\
    \midrule

    Shadow-Level (8:2) & 100   & 0    & 100   & 100   & 100   & 0 & 100 & 100 \\
    Shadow-Level (7:3) & 100   & 0    & 100   & 100   & 100   & 0 & 100 & 100 \\
    Shadow-Level (6:4) & 100   & 0.57 & 99.21 & 99.32 & 100   & 0 & 100 & 100 \\
    Shadow-Level (5:5) & 99.35 & 0    & 100   & 99.33 & 99.35 & 0 & 100 & 99.33 \\

    \midrule

    \multirow{2}{*}{\textbf{Method}} 
        & \multicolumn{4}{c}{MaliciousGen \& LMSYS-Chat} 
        & \multicolumn{4}{c}{MaliciousGen \& WildChat} \\
    \cmidrule(lr){2-5} \cmidrule(lr){6-9}
    
     & TPR & FPR & Precision & MCC & TPR & FPR & Precision & MCC \\
    \midrule

    Step-Level (8:2) & 97.14 & 16.74 & 63.85 & 71.07 & 98.57 & 9.35  & 76.24 & 82.22 \\
    Step-Level (7:3) & 99.00 & 7.75  & 86.46 & 88.55 & 100   & 9.75  & 83.68 & 86.90 \\
    Step-Level (6:4) & 99.20 & 7.43  & 90.51 & 90.83 & 100   & 15.86 & 82.78 & 83.95 \\
    Step-Level (5:5) & 99.03 & 14.14 & 88.22 & 85.95 & 99.68 & 18.28 & 85.36 & 83.15 \\
    \midrule

    Client-Level (8:2) & 100 & 13.04 & 70.00 & 78.02 & 100 & 6.52  & 82.35 & 87.74 \\
    Client-Level (7:3) & 100 & 7.50  & 86.96 & 89.69 & 100 & 5.00  & 90.91 & 92.93 \\
    Client-Level (6:4) & 100 & 11.43 & 86.21 & 87.38 & 100 & 8.57  & 89.29 & 90.35 \\
    Client-Level (5:5) & 100 & 17.24 & 86.11 & 84.42 & 100 & 13.79 & 88.57 & 87.38 \\
    \midrule

    Shadow-Level (8:2) & 100  & 0    & 100  & 100   & 100  & 0 & 100 & 100 \\
    Shadow-Level (7:3) & 100  & 0    & 100  & 100   & 100  & 0 & 100 & 100 \\
    Shadow-Level (6:4) & 99.60 & 0.57 & 99.20 & 98.97 & 99.60  & 0 & 100 & 99.66 \\
    Shadow-Level (5:5) & 100  & 0    & 100  & 100   & 100  & 0 & 100 & 100 \\
    
    \bottomrule
\end{tabular}
}
\caption{Evaluation results of the Llama3.1-8B-based malicious sample detector under varying malicious client ratios (TPR, FPR, Precision, and MCC).}
\label{tab:llama_detector}
\end{table*}

\subsection{Qwen2.5-7B under Varying Malicious Client Ratios}
\label{app:qwen_pre}
Results are shown in Table~\ref{tab:qwen_pre}.

\begin{table*}[t]
\centering
\resizebox{\textwidth}{!}{
\begin{tabular}{l cccc cccc}
    \toprule
    \multirow{2}{*}{\textbf{Method}} &  \multicolumn{4}{c}{BeaverTails \& LMSYS-Chat} & \multicolumn{4}{c}{BeaverTails \& WildChat} \\
    \cmidrule(lr){2-5} \cmidrule(lr){6-9}
    
     & Rule & MD-Judge & RM & MT-1 & Rule & MD-Judge & RM & MT-1 \\
    \midrule

      FedAvg (10:0) & 97.69 & 96.15 & -0.88 & 5.14 & 97.31 & 95.00 & -0.79 & 5.37  \\
      FedAvg (8:2)  & 77.69 & 66.92 & -2.05 & 5.06 & 67.12 & 51.92 & -1.96 & 5.05  \\
      FedAvg (7:3)  & 60.77 & 44.23 & -3.07 & 5.15 & 61.35 & 37.69 & -2.49 & 5.47  \\
      FedAvg (6:4)  & 55.00 & 39.81 & -3.26 & 4.97 & 49.42 & 23.46 & -3.14 & 5.26  \\
      FedAvg (5:5)  & 39.04 & 24.42 & -4.02 & 5.17 & 39.62 & 14.04 & -3.69 & 5.59  \\
    \midrule

    \multirow{2}{*}{\textbf{Method}} &  \multicolumn{4}{c}{MaliciousGen \& LMSYS-Chat} & \multicolumn{4}{c}{MaliciousGen \& WildChat} \\
    \cmidrule(lr){2-5} \cmidrule(lr){6-9}
    
     & Rule & MD-Judge & RM & MT-1 & Rule & MD-Judge & RM & MT-1 \\
    \midrule

      FedAvg (10:0) & 97.50 & 96.73 & -0.83 & 5.17 & 97.69 & 95.77 & -0.76 & 5.27  \\
      FedAvg (8:2)  & 69.23 & 49.81 & -2.64 & 5.28 & 59.23 & 32.31 & -2.81 & 5.31  \\
      FedAvg (7:3)  & 59.42 & 31.35 & -3.49 & 4.93 & 55.58 & 26.35 & -3.22 & 5.52  \\
      FedAvg (6:4)  & 51.73 & 19.04 & -3.88 & 5.18 & 58.46 & 25.58 & -3.23 & 5.48  \\
      FedAvg (5:5)  & 48.27 & 13.65 & -4.24 & 5.22 & 49.62 & 15.38 & -3.74 & 5.36  \\
    \bottomrule
    
\end{tabular}
}
\caption{Evaluation results of Qwen2.5-7B under varying malicious client ratios.}
\label{tab:qwen_pre}
\end{table*}

\subsection{Qwen2.5-7B under Different Aggregation Methods}
\label{app:qwen_methods}
Results are shown in Table~\ref{tab:qwen_methods}.

\begin{table*}[t]
\centering
\resizebox{\textwidth}{!}{
\begin{tabular}{l cccc cccc}
    \toprule
    \multirow{2}{*}{\textbf{Method}} &  \multicolumn{4}{c}{BeaverTails \& LMSYS-Chat} & \multicolumn{4}{c}{BeaverTails \& WildChat} \\
    \cmidrule(lr){2-5} \cmidrule(lr){6-9}
    
    & Rule & MD-Judge & RM & MT-1 & Rule & MD-Judge & RM & MT-1 \\
    \midrule

    FedAvg       & 60.77 & 44.23 & -3.07 & 5.15 & 61.35 & 37.69 & -2.49 & 5.47 \\
    Multi-Krum   & 72.88 & 59.04 & -2.36 & 5.16 & 72.88 & 60.96 & -1.68 & 5.64 \\
    Trimmed Mean & 60.38 & 43.46 & -3.06 & 4.75 & 65.77 & 43.65 & -2.31 & 5.41 \\
    FoolsGold    & 72.12 & 60.96 & -2.24 & 5.16 & 72.88 & 57.88 & -1.82 & 5.27 \\
    Residual     & 60.77 & 45.00 & -3.03 & 5.15 & 63.65 & 41.15 & -2.34 & 5.23 \\
    DnC          & 80.58 & 67.12 & -1.92 & 4.58 & 76.54 & 64.62 & -1.56 & 5.28 \\
    FLAME        & 77.88 & 61.15 & -2.18 & 4.45 & 75.00 & 62.31 & -1.57 & 5.07 \\
    LASA         & 67.50 & 49.81 & -2.72 & 4.47 & 68.46 & 46.73 & -2.18 & 5.14 \\
    \textbf{Step-Level}   & 70.38 & 58.46 & -2.42 & 5.04 & 83.65 & 72.12 & -1.34 & 5.13 \\
    \textbf{Client-Level} & 69.04 & 56.73 & -2.52 & 5.14 & 83.85 & 75.77 & -1.15 & 5.49 \\
    \textbf{Shadow-Level} & \underline{\textbf{97.88}} & \underline{\textbf{96.92}} & \underline{\textbf{-0.87}} & 5.17 & \underline{\textbf{97.50}} & \underline{\textbf{95.96}} & \underline{\textbf{-0.80}} & 5.39 \\

    \midrule

    \multirow{2}{*}{\textbf{Method}}& \multicolumn{4}{c}{MaliciousGen \& LMSYS-Chat} & \multicolumn{4}{c}{MaliciousGen \& WildChat} \\
    \cmidrule(lr){2-5} \cmidrule(lr){6-9}
    
    & Rule & MD-Judge & RM & MT-1 & Rule & MD-Judge & RM & MT-1 \\
    \midrule
    FedAvg       & 59.42 & 31.35 & -3.49 & 4.93 & 55.58 & 26.35 & -3.22 & 5.52 \\
    Multi-Krum   & 72.69 & 60.58 & -2.19 & 5.20 & 72.12 & 56.15 & -1.93 & 5.24 \\
    Trimmed Mean & 68.85 & 48.85 & -2.73 & 5.28 & 56.35 & 30.38 & -2.99 & 5.43 \\
    FoolsGold    & 55.96 & 23.85 & -3.85 & 5.28 & 65.00 & 40.77 & -2.65 & 5.61 \\
    Residual     & 57.69 & 30.77 & -3.50 & 5.08 & 54.23 & 22.88 & -3.30 & 5.36 \\
    DnC          & 76.54 & 63.27 & -2.17 & 4.75 & 73.08 & 60.58 & -1.91 & 5.19 \\
    FLAME        & 79.23 & 69.23 & -1.85 & 4.79 & 73.85 & 62.31 & -1.74 & 5.08 \\
    LASA         & 68.27 & 50.77 & -2.69 & 4.81 & 55.96 & 27.69 & -3.27 & 5.30 \\
    \textbf{Step-Level}   & 93.08 & 92.12 & -1.05 & 5.16 & \underline{\textbf{97.69}} & 95.38 & -0.80 & 5.24 \\
    \textbf{Client-Level} & 83.27 & 73.27 & -1.71 & 5.01 & 97.31 & 95.77 & \underline{\textbf{-0.76}} & 5.33 \\
    \textbf{Shadow-Level} & \underline{\textbf{97.31}} & \underline{\textbf{96.35}} & \underline{\textbf{-0.92}} & 5.06 & \underline{\textbf{97.69}} & \underline{\textbf{96.35}} & -0.77 & 5.68 \\
    \bottomrule
    
\end{tabular}
}
\caption{Evaluation results of Qwen2.5-7B with a 30\% malicious client ratio across different aggregation methods.}
\label{tab:qwen_methods}
\end{table*}

\end{document}